\documentclass[12pt]{spieman}  
\usepackage{amsmath,amsfonts,amssymb}
\usepackage{graphicx}
\usepackage{setspace}
\usepackage{tocloft}

\title{Resonant helical dichroism in twisted dielectric metastructures}

\author[a,b,c,$\dagger$]{Yiyuan~Wang}
\author[d,$\dagger$]{Chi~Li}
\author[d]{Haoyi~Yu}
\author[d,e]{Stefan~A.~Maier}
\author[a,c,*]{Jinhui~Shi}
\author[d,*]{Haoran~Ren}
\author[b,*]{Kirill~Koshelev}
\affil[a]{Key Laboratory of In-Fiber Integrated Optics of Ministry of Education, College of Physics and Optoelectronic Engineering, Harbin Engineering University, Harbin 150001, China}
\affil[b]{Research School of Physics, Australian National University, Canberra, ACT 2601, Australia}
\affil[c]{Key Laboratory of Photonic Materials and Devices Physics for Oceanic Applications, College of Physics and Optoelectronic Engineering, Harbin Engineering University, Harbin 150001, China}
\affil[d]{School of Physics and Astronomy, Faculty of Science, Monash University, Melbourne, Victoria 3800, Australia}
\affil[e]{Department of Physics, Imperial College London SW7 2AZ, United Kingdom}
\affil[$\dagger$]{These authors contributed equally to this work}

\cftpagenumbersoff{figure}
\cftpagenumbersoff{table} 
\begin{document} 
\maketitle

\begin{abstract}
Circular dichroism, arising from interactions with light fields of opposite spin angular momentum, has become a fundamental tool for molecular characterization. Meanwhile, helical dichroism (HD)—the dichroic response to vortex beams carrying opposite orbital angular momentum (OAM)—offers an alternative approach for probing chiral molecules and photonic structures. Previous demonstrations of HD have been limited to non-resonant light-matter interactions with chiral micro- and nanostructures, leaving the realization of resonance helical dichroism largely unexplored. Here, we present the design and implementation of twisted dielectric metastructures, composed of an array of rotated silicon trimer nanostructures harnessing nonlocal photonic modes with a high quality factor of several dozen that enable strong resonant HD for OAM values up to $10$. We experimentally demonstrate resonantly enhanced HD for strongly focused OAM beams with the magnitude of topological charges from $1$ to $3$. Our findings pave the way for resonant nanophotonics involving OAM beams, unlocking the full potential of structured light for applications in molecular sensing, optical imaging, nonlinear optics, and optical data storage.      
\end{abstract}


{\noindent \footnotesize\textbf{*} Jinhui Shi~\linkable{shijinhui@hrbeu.edu.cn}; Haoran Ren~\linkable{haoran.ren@monash.edu}; Kirill Koshelev~\linkable{kirill.koshelev@anu.edu.au}}

\begin{spacing}{2}   

\section{Introduction}
\label{sect:intro}  

Structured light involves the spatial and temporal shaping of light in multiple degrees of freedom, including its amplitude, phase, polarization, and angular momentum. The spin angular momentum (SAM) of light manifests in left-handed and right-handed circularly polarized (LCP and RCP) beams, corresponding to the spin values $\sigma = \pm 1$. In addition to SAM, light can also carry orbital angular momentum (OAM), which is associated with a vortex wavefront and features a theoretically unbounded spectrum of orthogonal helical modes with distinct topological charges of integer $L$, ranging from -$\infty$ to +$\infty$. Owing to the unbounded vortex modes, OAM of light has garnered significant attention and enabled a wide range of photonic applications. These include optical trapping~\cite{la2004optical,stilgoe2022controlled}, imaging and data storage~\cite{fang2020orbital,lin2021spectral,gao2023metasurface}, holographic displays~\cite{ren2019metasurface,ren2020complex}, optical sensing~\cite{lavery2013detection,cheng2025metrology}, and quantum information processing~\cite{fickler2014interface,zhou2015quantum}.

\begin{figure}[t]
    \centering
    \includegraphics[width=0.75\linewidth]{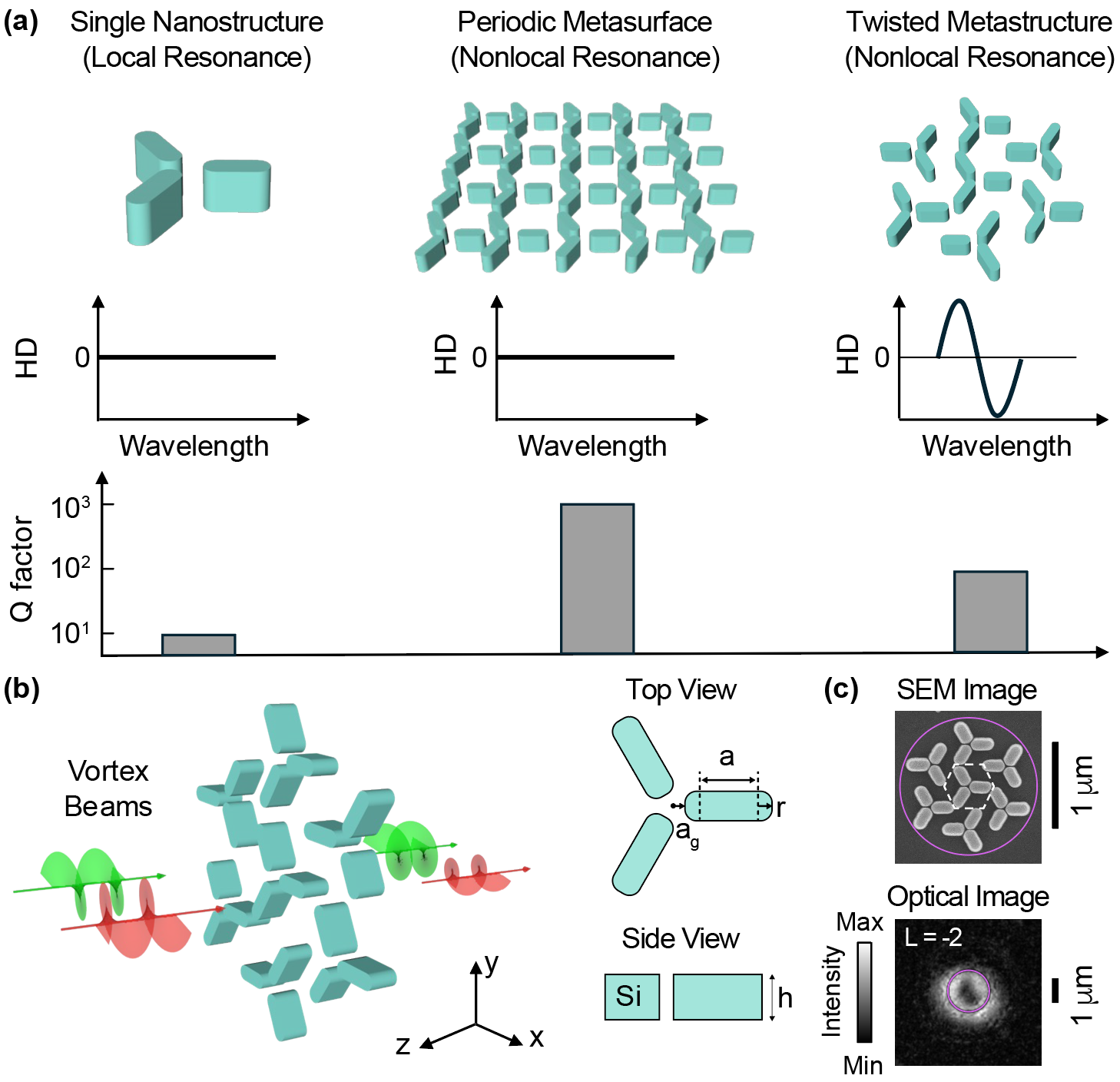}
    \caption{\label{fig:1} {\bf Concept of resonant helical dichroism in twisted dielectric metastructures.} (\textbf{a}) Local single trimer nanostructure with a low Q factor showing no HD, nonlocal periodic metasurface composed of periodic trimers with a very high Q factor showing no HD, and nonlocal twisted metastructures consisting of several trimers with a high Q factor showing a resonant HD. (\textbf{b}) (Left) Schematic of finite-size twisted Si metastructures illuminated with vortex beams carrying OAM with opposite signs that results into differential transmission inducing helical dichroic response. (Right) Si trimer unit cell with geometric parameters. (\textbf{c}) SEM image of unit cell and optical image of beam with $L=-2$. Hexagonal unit cell is shown with a white dashed line. Scale bar is $1$~$\mu$m. The purple circle indicates the 7-unit metastructure footprint.}       
\end{figure}

Circular dichroism (CD), a technique that measures the differential absorption or transmission of LCP and RCP light beams as they pass through a chiral medium without mirror symmetry, has become an essential tool for molecular characterization in the biochemistry and pharmaceutical industries~\cite{berova2000circular,lewis2005dna,rogers2019electronic}. Recently, significant advancements in chiroptical interactions have been achieved using resonant chiral structures, enabling the resonant enhancement of CD signals with high quality factors ($Q$
factors)~\cite{koshelev2018asymmetric,gorkunov2020metasurfaces,shi2022planar}. This resonant CD effect facilitates highly efficient interactions between light and matter, allowing for a wide range of phenomena and applications~\cite{zhao2017chirality,seo2021circularly,zhang2022chiral,koshelev2023resonant}. Conversely, helical dichroism (HD), also called vortical differential scattering, is an related optical phenomenon that measures the differential absorption or scattering of light with opposite OAM states as they interact with a medium. HD extends the capabilities of traditional chiral spectroscopy by leveraging OAM-based light-matter interactions, making it valuable in both fundamental research and practical applications~\cite{dai2023robust}. However, previous HD demonstrations have utilized non-resonant chiral micro- and nanostructures~\cite{ni2021gigantic,ni2021giant,liu2022tailoring}, focusing on characterization at a single wavelength, as schematically shown in Fig.~\ref{fig:1}(a). To date, resonant HD with strong light-matter interactions between vortex beams and resonant chiral structures remains unexplored.

In this Letter, we propose, design and test experimentally metastructures to explore and demonstrate resonant HD marked with the excitation of dark modes. The concept of dark modes is closely associated with the phenomenon of optical bound states in the continuum (BICs) receiving much attention in the last decade~\cite{hsu2016bound,koshelev2019meta,koshelev2023bound}. Figure~\ref{fig:1}(a) demonstrates the concept of twisted finite-size metastructures with a high-quality resonance that enable a resonant helical dichroic response via the simultaneous fulfillment of three conditions: (i) the laser wavelength matches the resonant wavelength, (ii) the OAM beam spot matches the structure footprint, and (iii) geometric twist breaks the mirror symmetries for nonzero helical dichroism but preserves the rotational symmetry, required for observation of dark modes. In comparison, single nanostructures could only satisfy the first two criteria and support low-quality modes. On the other hand, periodic metasurfaces do not satisfy the last two criteria despite their ability to provide an extremely sharp optical response.

Our design is based on a finite-size hexagonal array of twisted silicon (Si) trimers with a small number of unit cells, see Fig.~\ref{fig:1}(b). We start with designing infinitely large periodic metasurfaces without twist that support high-quality dark modes in the near-IR wavelength range. We then track the dark mode evolution in finite-size structures in dependence on the number of unit cells and the degree of unit cell twist. We refer to such finite-size metasurfaces as {\it metastructures}. We numerically calculate near- and far-field characteristics of $7$-unit twisted metastructures, scattering cross-section and local field enhancement, respectively, under illumination with strongly focused Laguerre-Gaussian vortex beams. We show that both in the near- and far-field the HD is resonantly enhanced in the vicinity of the dark mode wavelength. We fabricate the designed metastructures and test their HD response experimentally, see Fig.~\ref{fig:1}(c). The demonstrated HD in the scattering far-field range shows resonant features, confirming the proposed concept. 

\section{Results}

\subsection{Engineering dark quasi-BICs with dichroic mode structure}

As the first step, we design an infinite periodic metasurface supporting a dark mode that consists of a hexagonal lattice of Si trimers suspended in air, shown in the right part of Fig.~\ref{fig:1}(b). The unit cell parameters are $a=150$~nm, $a_{\rm g}=25$~nm, $r=40$~nm, period is $250$~nm, thickness $h$ is $140$~nm. The point group symmetry of the unit cell is the cyclic group with three vertical mirror planes, $C_{3v}$, which has two non-degenerate irreducible representations corresponding to non-radiating dark BIC modes~\cite{gelessus1995multipoles}. We use eigenmode solver in COMSOL Multiphysics (see Methods in SM) to show that the designed structure supports such a dark mode in the near-infrared (near-IR) range at $692$~nm with a nonzero out-of-plane $E_z$ component of the electric field that is not radiating to the far field (see Fig.S4 in SM). 
\begin{figure}[t]
    \centering
    \includegraphics[width=0.5\linewidth]{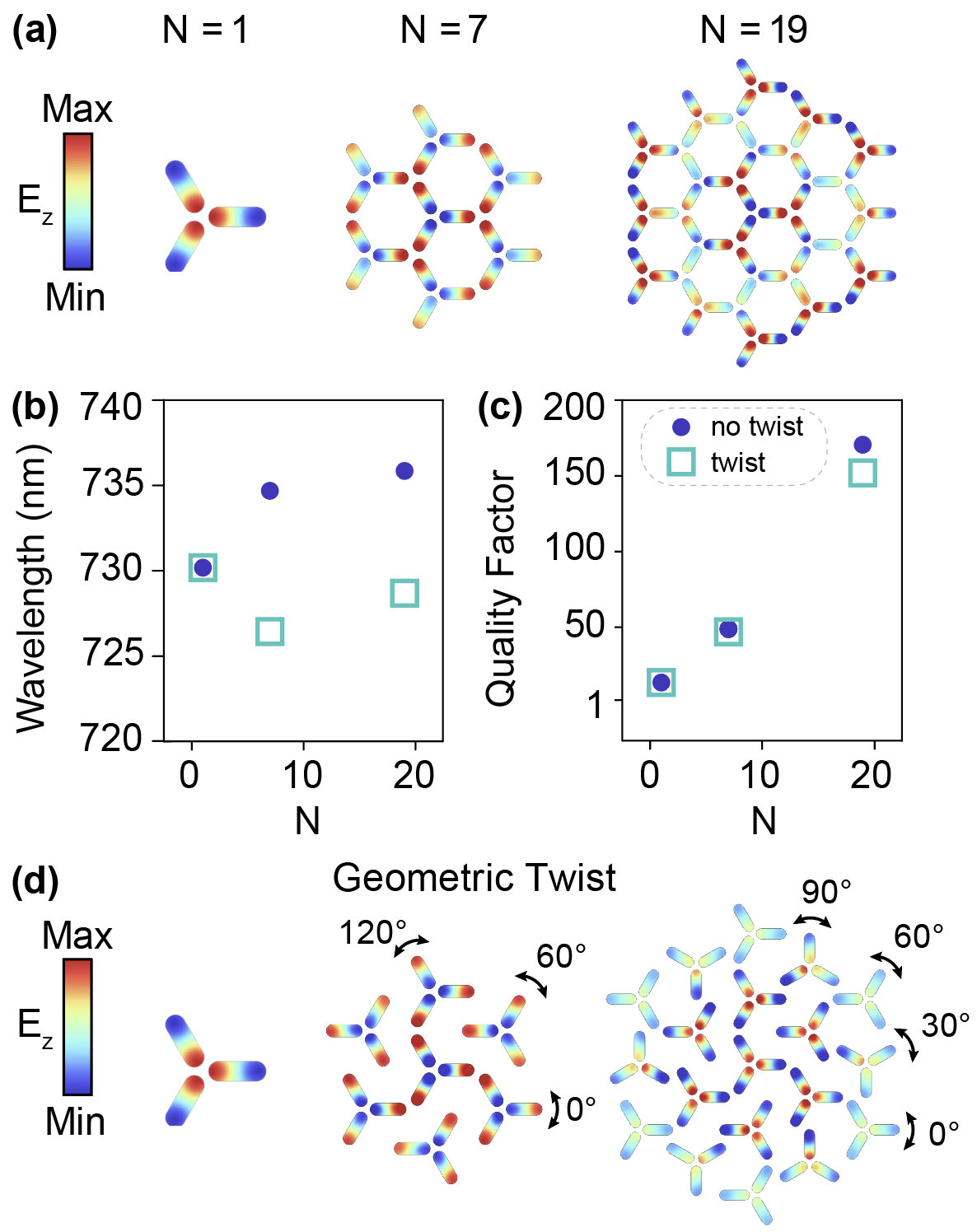}
    \caption{\label{fig:2} {\bf Dark modes in finite-size metastructures.} (\textbf{a}) Near-field distribution of the out-of-plane electric field $E_z$ of dark quasi-BIC modes depending on the number of unit cells $N$ from $1$ to $19$. (\textbf{b,c}) The evolution of the resonant wavelength (\textbf{b}) and Q factor (\textbf{c}) of the dark mode with the increase of $N$ for metastructures without twist (blue circle) and with twist (cyan open square). (\textbf{d}) Effect of the twist on the near-field profiles of dark modes. The examples twist angles are shown with pictograms.}
\end{figure}

Excitation of conventional metasurfaces with structured OAM beams is inefficient because the beam only covers a small doughnut area and has an intensity minimum in the center, as shown in the experimental image of an OAM beam overlapping with 7-unit twisted metastructures in Fig.~\ref{fig:1}(c). Therefore, we engineer finite-size metastructures supporting dark modes of similar nature as BICs following the design parameters for an infinite structure.  

Figure~\ref{fig:2}(a) shows the out-of-plane $E_z$ field component for the dark modes realized in metastructures with the total number of unit cells $N=1,\ 7,\ 19$. Here, $N=7$ corresponds to one layer of the nearest neighbors of meta-molecules, and $N=19$ corresponds to two layers. The dark mode field profiles closely resemble the corresponding BIC mode of the infinite metasurface with distortions at the interfaces due to edge effects. Figures~\ref{fig:2}(b,c) show the dependence of resonant mode wavelength and Q factor on $N$ (blue dots). The designed modes do not radiate normally to the surface plane but can radiate to oblique angles due to finite size effects. This explains the finite value of Q factor increasing with $N$, following the growth rate characteristic to BIC modes~\cite{bulgakov2017light,zakomirnyi2019collective}.

The periodic arrangements of trimers in finite-size metastructures, shown in Fig.\ref{fig:2}(a), is associated with in-plane mirror symmetries that prevent dichroic scattering of OAM vortex beams. We break the mirror symmetries by introducing a new concept of {\it geometric twist}, shown in Fig.\ref{fig:2}(d), where each unit cell in the first layer of neighbors is rotated by an integer of $60$ degrees, and in the second layer of neighbors - by $30$ degrees, which adds up to full $2\pi$ rotation over the circumference. The twist does not distort the third-order rotational symmetry that protects the dark mode, thus the electric field profiles look similar to the non-twisted case, with less energy localized at the structure edges.  Figures~\ref{fig:2}(b,c) show that after twisting, the mode wavelengths blue shift by $10$~nm and the Q factor does not change much.

\subsection{Near-field analysis of resonant helical dichroic response}

\begin{figure}[t]
    \centering
    \includegraphics[width=0.65\linewidth]{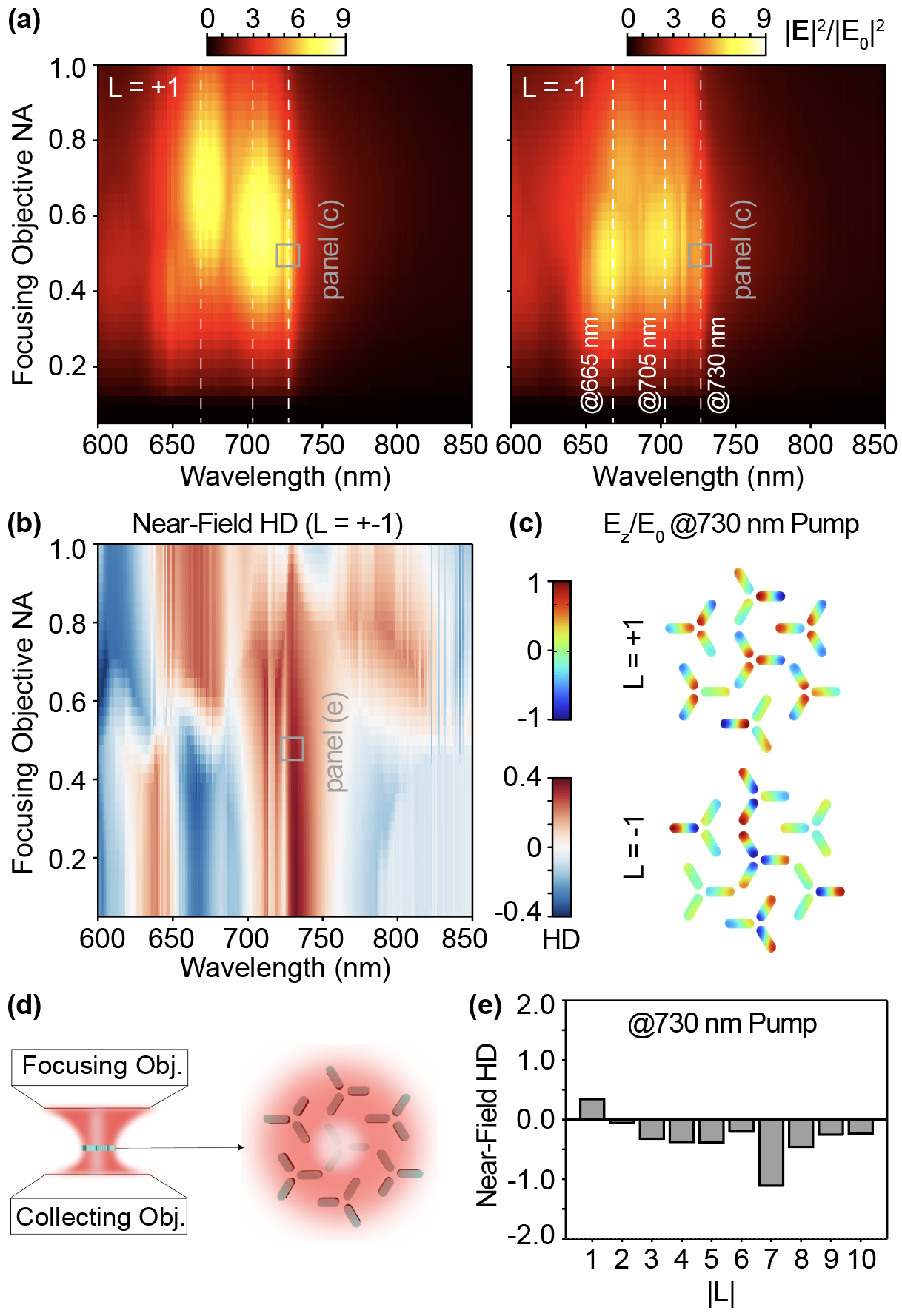}
    \caption{\label{fig:3} {\bf Calculated near-field enhancement and helical dichroism for a $7$-unit metastructure.} (\textbf{a},\textbf{b}) Local field enhancement $u$ (\textbf{a})  given by Eq.~\eqref{eq:0} and near-field HD (\textbf{b})  given by Eq.~\eqref{eq:1} vs. wavelength and NA of focusing objective for $L=\pm 1$. The white dashed lines show three resonant regimes at $665$, $705$ and $730$~nm, respectively. (\textbf{c}) Near-field profile $E_z/E_0$ at $730$~nm with the  focusing NA of $0.5$. (\textbf{d}) Schematic of focusing and collection with a doughnut beam shape in the structure plane. (\textbf{e}) Resonant near-field HD at $730$~nm vs. $|L|$ for focusing NA $0.5$. }
\end{figure}

Next, we analyze how the designed finite-size metastructures respond to excitation with structured beams carrying different OAM in the near-field. We derive an exact expression for the transversal and longitudinal field components of a linearly polarized Laguerre-Gaussian beams with the OAM value of $L$ and the radial index $0$ in an isotropic environment that allows studying highly non-paraxial regime (see Methods in SM). The beam focus spot is modeled via the NA of focusing objective, as shown in Fig.~\ref{fig:3}(d). We use the derived expressions as a background field in a frequency domain solver in COMSOL Multiphysics (see Methods in SM) and calculate the local field enhancement $u$ defined as~\cite{maier2007plasmonics}
\begin{equation}
u=\frac{|\mathbf{E}|^2}{|\rm E_0|^2},    
\label{eq:0}
\end{equation}
where $|\mathbf{E}|$ is the maximal electric field value within the Si volume, and $E_0$ is the pump field amplitude. 

Figure~\ref{fig:3}(a) shows the dependence of the local field enhancement on wavelength and NA of the focusing objective for $L=\pm 1$ in the visible and near-IR wavelength range for the 7-unit metastructure with the parameters as for Fig.~\ref{fig:2}. There are three resonant regimes providing an enhancement of the local fields, at $665$~nm, $705$~nm and $730$~nm, respectively for NA in the range from $0.3$ to $0.8$. The degree of enhancement for opposite OAM values is similar at $665$~nm and $705$~nm,  but drastically different at $730$~nm. Figure~\ref{fig:3}(c) shows the near-field distribution of the out-of-plane electric field component at $730$~nm and NA $0.5$, with the profile and wavelength matching the designed BIC dark mode for a 7-unit structure in Fig.~\ref{fig:2}(d). 

We then evaluate the near-field HD via the local field enhancement $u$ for OAM beams with opposite signs as
\begin{equation}
{\rm HD}_{\rm NF} = 2\frac{u_{|L|}-u_{-|L|}}{u_{|L|}-u_{-|L|}}.
\label{eq:1}
\end{equation}
Figure~\ref{fig:3}(b) shows that the highest resonant enhancement of near-field HD is achieved in the narrow vicinity of $730$~nm. The maximal value of ${\rm HD}_{\rm NF}$ is $0.4$. We next study the resonant response at $730$~nm for $|L|$ from $1$ to $10$ for NA of $0.5$, shown in Fig.~\ref{fig:3}(e). The near-field HD has strong resonant behavior at the wavelength of the designed dark BIC for $|L|=1,3,4,5,7,8$ with the maximal value ${\rm HD}_{\rm NF}\simeq -1.1$ for $|L|=7$.

\subsection{Experimental verification of resonant helical dichroism}
\begin{figure}[t]
    \centering
    \includegraphics[width=0.9\linewidth]{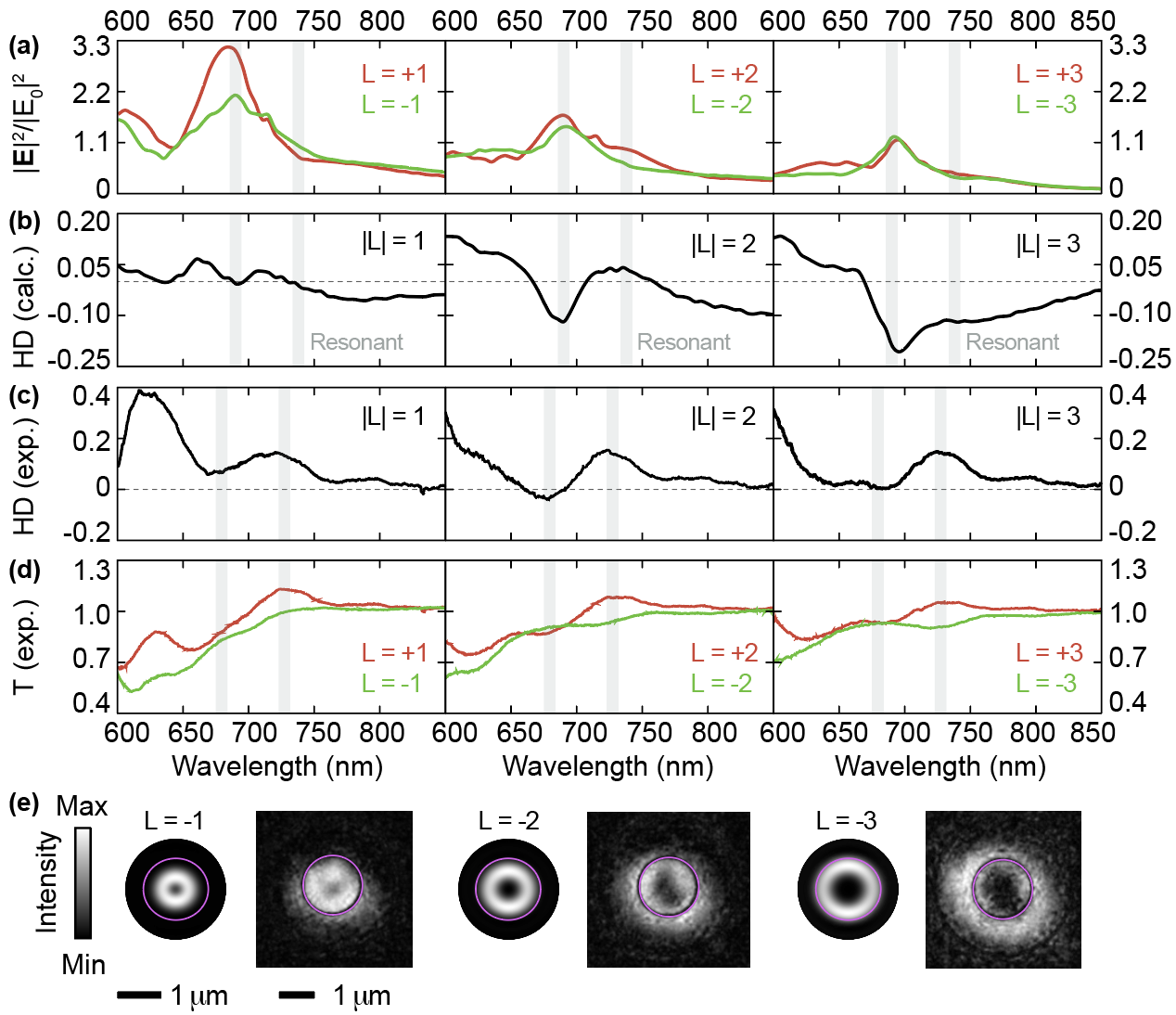}
    \caption{\label{fig:4} \textbf{Experimental observation of resonant HD for a 7-unit metastructure and comparison with calculations.}  (\textbf{a}) The local field enhancement $u$ calculated via Eq.~\eqref{eq:0} for $n_{\rm s}=1.4$, focusing NA $1.40$, collecting NA $0.9$, and sample thickness $h=140$~nm .  (\textbf{b}) HD extracted from the calculated transmittance using Eq.~\eqref{eq:2} for the same parameters as in (\textbf{a}). (\textbf{c}) HD extracted from the measured transmittance using Eq.~\eqref{eq:2}. (\textbf{d}) Measured transmittance for $L=\pm1,\pm2,\pm3$.  The gray shaded areas in all panels show the resonant regimes. (\textbf{e}) The calculated at $700$~nm (left) and measured at  $650$~nm (right) images of the pump beam profiles for $L=-1,-2,-3$. Purple circles indicate the footprint area of the 7-unit sample.}
\end{figure}

Next, we evaluate how the designed helical dichroic response can be observed experimentally in the transmission far-field response. We fabricate a set of $7$- and $19$-unit twisted metastructures made of amorphous Si on a quartz substrate, following the design in Figs.~\ref{fig:1} and~\ref{fig:2} (see Methods and Fig.S1 in SM). The geometrical parameters are the same as the calculations above, $a=150$~nm, $a_{\rm g}=25$~nm, $r=40$~nm, period is $250$~nm, $h=140$~nm. Thickness of the resulting sample is confirmed with atomic force microscope measurements (see Fig.S2 in SM). 

We create linearly polarized Laguerre-Gaussian beams with $|L|$ from $1$ to $5$ with spatial light modulators with the central wavelength of $620$~nm. We use a collecting objective with the NA of $0.9$ and an oil immersion objective with the NA of $1.45$ to focus the OAM beams (see Methods in SM). The focused beam profiles are shown in Figure~\ref{fig:4}(e) for $L=-1,-2,-3$ on the right side of the corresponding column (see Fig.S3 in SM for beam profiles with a larger $|L|$).

We first numerically analyze the near- and far-field response  of the 7-unit twisted metastructure in the experimental conditions. We model the substrate and oil as an isotropic media with the refractive index $n_{\rm s}=1.4$. The other geometrical parameters are as in the initial design, and the focusing NA is $1.40$. The spatial profiles of focused OAM beams are shown in Figure~\ref{fig:4}(e) for $L=-1,-2,-3$ on the left side of the corresponding column.

Figure~\ref{fig:4}(a) shows the calculated near-field enhancement $u$ for $L = \pm1,\pm2,\pm3$ defined as in Eq.~\eqref{eq:0}. The peaks of $u$ at $680$ and $735$~nm are shown with gray shaded areas. We next calculate the transmittance $T$ as the forward scattered power through the collecting aperture normalized on the beam power through the same aperture. The collection NA is $0.9$ as in the experiment. We evaluate the transmission HD via transmittance of OAM beams with opposite signs
\begin{equation}
{\rm HD} = 2\frac{T_{|L|}-T_{-|L|}}{T_{|L|}+T_{-|L|}}.
\label{eq:2}
\end{equation} 
Figure~\ref{fig:4}(b) shows the calculated far-field HD spectra for $|L|=1,2,3$. The HD lineshape shows dip and peak features at the corresponding resonant wavelengths.

We next measure the transmittance spectra $T$ of the $7$-unit metastructure in wavelength range from $600$ to $850$~nm. The tranmittance is defined as the ratio of power scattered by metastructure in the forward direction and the respective forward scattered power from the quartz substrate. The experimental HD, shown in Fig.~\ref{fig:4}(c), is evaluated from the transmittance data using Eq.~\eqref{eq:2}. The spectra show direct correspondence to the numerical results in Fig.~\ref{fig:4}(b) with larger absolute values of HD. By comparing the experiment and simulation results, we deduct the resonant positions in the experimental data, marked with gray shaded areas. The experimental data exhibits a blue shift of $15$~nm. The corresponding resonant features can be visually seen in the experimental transmittance spectra in Fig.~\ref{fig:4}(d).

\section{Discussion and conclusions}

A high degree of helical dichroic response observed in the local field properties in Fig.~\ref{fig:3} can be associated with the spatial structure of the dark mode that matches the phase profile of helical beams. The overlap of mode and beam profiles is enhanced even more for higher values of $|L|$, as can be seen from Fig.~\ref{fig:3}(e). 

The observed resonant properties in the experimental sample in Fig.~\ref{fig:4}(c) are blue-shifted compared to the simulation data in Fig.~\ref{fig:4}(b). We associate such blue shift with the following reasons. Firstly, the material absorption is larger than for normal amorphous silicon due to potential limitations of ellipsometry measurements (see Fig.S1 in SM), which could affect the mode position and HD profile. Secondly, the central wavelength of spatial light modulators is $620$~nm, thus a minor distortion of Laguarre-Gaussian form can occur at larger wavelengths due to dispersion effects. Thirdly, the fabricated samples show moderately uneven thickness profile and rounded edges (see Fig.S2), which could further affect the mode wavelength. Lastly, the experimental setup features a quartz substrate and the immersion oil used to symmetrize the dielectric environment, that is required for the observation of dark modes~\cite{hsu2013observation}. The simulation models the substrate and immersion oil as an isotropic environment with a refractive index of $1.4$.

To conclude, we have proposed and verified a concept of resonant helical dichroism allowing for the use of resonant metaphotonics platform to enhance light-matter interactions with helical beams carrying OAM. We have designed a planar subwavelength metastructure that supports resonances with a high quality factor in the near-infrared range. We have introduced a concept of geometrical twist by breaking mirror symmetries but preserving rotational symmetries that allows helical dichroic response via matching the phase profiles of twisted light beams. We fabricated designed twisted metastructures and verified the concept of resonant HD in far-field optical experiments. The measured data evidences resonant helical dichroic response in the visible range. Our twisted metastructures, designed to support resonant helical dichroism, have the potential to fully harness structured light for transformative applications in molecular sensing, optical imaging, nonlinear optics, and high-density optical data storage.

\subsection*{Disclosures}
No conflicts of interest, financial or otherwise, are declared by the authors.

\subsection*{Code, Data, and Materials Availability} 
The data and code that support the findings of this study are available from the corresponding authors upon request.

\subsection*{Author Contributions}
Y.W., H.R. and K.K. conceived the project. Y.W. and K.K. developed the theoretical model and performed numerical simulations. C.L. fabricated the samples. C.L., H.Y. and H.R. performed the measurements. Y.W., C.L. and K.K. processed the experimental data. Y.W., C.L., H.R. and K.K. wrote the manuscript with input from all authors. S.A.M., and J.S. participated in discussions. H.R. and K.K. supervised the project.

\subsection* {Acknowledgments}
J.S. acknowledges funding supported by the National Natural Science Foundation of China (62275061) and Fundamental Research Funds for the Central Universities (3072024LJ2502, 3072025YC2502). H. R. and S. A. M. acknowledge funding support from the Australian Research Council (DP220102152). H. R. acknowledges the DECRA (DE220101085) funding support from the Australian Research Council. This work was performed in part at the Melbourne Centre for Nanofabrication (MCN) in the Victorian Node of the Australian National Fabrication Facility (ANFF). K.K. acknowledges the DECRA (DE250100419) funding support from the Australian Research Council.


\bibliography{bibliography}   
\bibliographystyle{spiejour}   





\end{spacing}
\end{document}